\documentclass[conference]{IEEEtran}
\IEEEoverridecommandlockouts

\usepackage{cite}
\usepackage{amsmath,amssymb,amsfonts}
\usepackage{algorithmic}
\usepackage{graphicx}
\usepackage{textcomp}
\usepackage{xcolor}
\usepackage{comment}
\usepackage{multirow}
\usepackage{float}
\usepackage{placeins}
\usepackage{url}
\usepackage{booktabs}
\usepackage{tabularx} 
\usepackage{enumitem}
\usepackage{tikz}
\usepackage{dblfloatfix} 

\def\BibTeX{{\rm B\kern-.05em{\sc i\kern-.025em b}\kern-.08em
    T\kern-.1667em\lower.7ex\hbox{E}\kern-.125emX}}
\begin{document}

\title{From Theory to Practice: Code Generation Using LLMs for CAPEC and CWE Frameworks\\

\thanks{This work was partially funded by Office of Naval Research (ONR).}
}

\author{
    Murtuza Shahzad, Joseph Wilson, Ibrahim Al Azher, Hamed Alhoori, Mona Rahimi \\ 
    \{msyed1, jwilson33, iazher1, alhoori, rahimi\}@niu.edu\\
    Department of Computer Science \\ 
    Northern Illinois University \\
    DeKalb, Illinois USA \\
    \vspace{-20pt}
}

\maketitle

\begin{abstract}
The increasing complexity and volume of software systems have heightened the importance of identifying and mitigating security vulnerabilities. The existing software vulnerability datasets frequently fall short in providing comprehensive, detailed code snippets explicitly linked to specific vulnerability descriptions, reducing their utility for advanced research and hindering efforts to develop a deeper understanding of security vulnerabilities. To address this challenge, we present a novel dataset that provides examples of vulnerable code snippets corresponding to Common Attack Pattern Enumerations and Classifications (CAPEC) and Common Weakness Enumeration (CWE) descriptions. By employing the capabilities of Generative Pre-trained Transformer (GPT) models, we have developed a robust methodology for generating these examples. Our approach utilizes GPT-4o, Llama and Claude models to generate code snippets that exhibit specific vulnerabilities as described in CAPEC and CWE documentation. This dataset not only enhances the understanding of security vulnerabilities in code but also serves as a valuable resource for training machine learning models focused on automatic vulnerability detection and remediation. Preliminary evaluations suggest that the dataset generated by Large Language Models demonstrates high accuracy and can serve as a reliable reference for vulnerability identification systems. We found consistent results across the three models, with 0.98 cosine similarity among codes. The final dataset comprises 615 CAPEC code snippets in three programming languages: Java, Python, and JavaScript, making it one of the most extensive and diverse resources in this domain. This research contributes to the field of cybersecurity by introducing an innovative dataset that supports advanced studies on software vulnerabilities and facilitates the development of tools for their prevention and mitigation.
\end{abstract}


\section{Introduction}
Software security is a critical concern in the contemporary digital landscape, where the integrity and reliability of applications are paramount. Cybercrimes are estimated to cost \$10.5 trillion annually worldwide by 2025\cite{sharif2022literature}. As software systems continue to grow in complexity, the task of detecting and addressing security vulnerabilities becomes more challenging. Vulnerabilities in software can lead to severe consequences, including data breaches, unauthorized access, and significant financial losses \cite{romanosky2016examining, meisner2017financial}. Hence, there is a pressing need for effective tools and methodologies to detect and understand these vulnerabilities early in the software development lifecycle.

Common Weakness Enumeration (CWE) and Common Attack Pattern Enumeration and Classification (CAPEC) are key frameworks provided by MITRE\footnote{https://www.mitre.org/}, a nonprofit organization that operates federally funded research and development centers (FFRDCs), for cataloging and describing software weaknesses and vulnerabilities. CAPEC provides a comprehensive catalog describing the attack techniques and methodologies used by adversaries, aiding in threat modeling and defense strategy development. CWE lists software and hardware weaknesses that can lead to vulnerabilities, helping developers and security professionals identify and mitigate security flaws. Together, CAPEC and CWE enhance the understanding of potential threats and system weaknesses to improve cybersecurity measures. The CAPEC framework is instrumental in guiding risk-based security testing by enabling the identification and prioritization of attack patterns \cite{Seehusen2015}. All the CAPEC and CWE have their unique identifiers (IDs). For each CAPEC ID, the CAPEC data\footnote{https://capec.mitre.org/data/} provides the related CWE IDs from CWE data\footnote{https://cwe.mitre.org/data/}. In most cases, there is a one-to-many relationship between CAPEC and CWE. For example, \textit{CAPEC-66: SQL Injection} is associated with two related weaknesses, \textit{CWE-89: Improper Neutralization of Special Elements used in an SQL Command (`SQL Injection')} and \textit{CWE-1286: Improper Validation of Syntactic Correctness of Input}. In some cases, the CAPEC does not have any related CWEs. For example, \textit{CAPEC-165: File Manipulation} has no related CWEs.  

Despite a comprehensive catalog provided by CAPEC and CWE, there remains a gap in practical resources that illustrate how these vulnerabilities manifest in actual code. There is a substantial deficiency in practical code examples that programmers can utilize to better comprehend and recognize the attack patterns and weaknesses prevalent in today’s cybersecurity landscape. We explored the latest version of CAPEC (version 3.9) and CWE (version 4.14) and found that a significant number of CAPECs (\( \approx 40\% \)) and CWEs (\( \approx 47\% \)) lack corresponding code examples as shown in Table \ref{tab:stats}. We also provide statistics on current code availability for Java, JavaScript, and Python languages, highlighting that the availability is particularly low for JavaScript and Python. 

The availability of practical code examples is crucial for bridging the gap between theoretical descriptions of vulnerabilities and their real-world manifestations. These examples would serve as a valuable resource for developers, security researchers, and educators, enabling them to better understand the mechanics of vulnerabilities and attack patterns. For developers, having access to code examples allows them to recognize potential flaws in their own applications, thereby facilitating the implementation of secure coding practices. Security researchers can use these examples to design more effective detection tools and countermeasures tailored to specific vulnerabilities. Educators, in turn, can use these examples to create hands-on training materials that help students and professionals grasp the nuances of cybersecurity threats. Furthermore, code examples can aid in the development of automated vulnerability scanning tools by providing patterns to identify weaknesses in software systems. In essence, a rich repository of coding examples not only enhances the comprehension of CAPEC and CWE data but also strengthens the overall ecosystem of software security.

\begin{figure*}[h]
  \centering
  \includegraphics[width=0.80\textwidth]{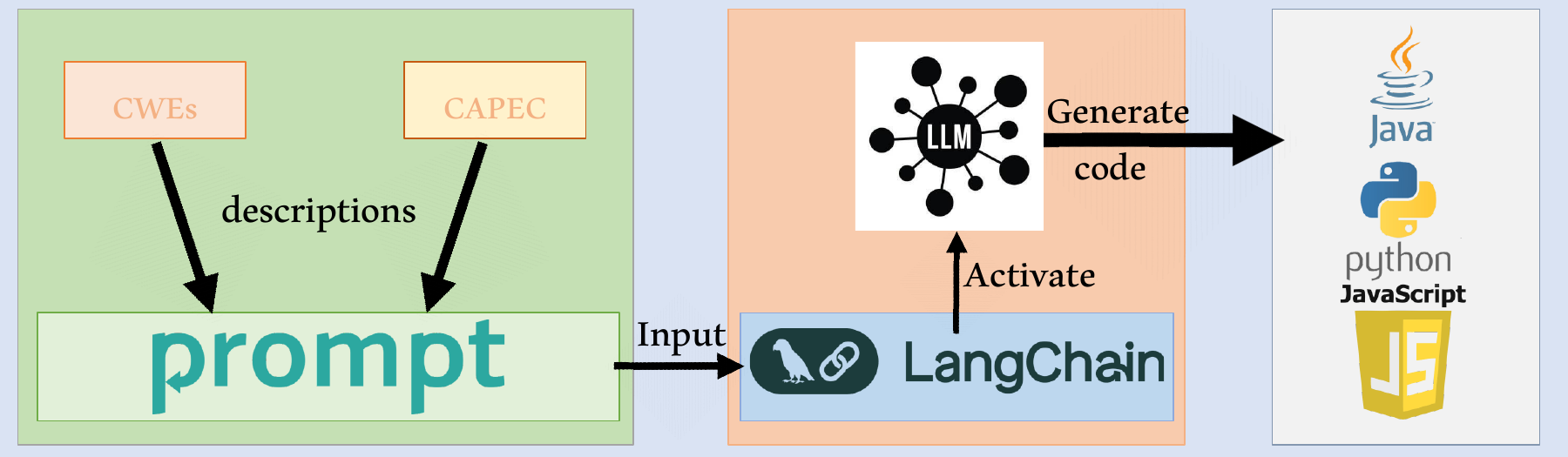}
  \caption{High-level Overview of Code Generation Process.}
  \label{fig:diagram}
  \vspace{-15pt}
\end{figure*}

To address the need for practical code examples, we present a methodology for generating code for each CAPEC by leveraging the capabilities and contextual understanding of Large Language Models (LLMs). Figure \ref{fig:diagram} outlines the process of generating code for each CAPEC. The process consists of first retrieving the names and descriptions from CAPECs and their related CWEs. We then formulated a specific prompt, tailored to our requirements, and applied it using LangChain\cite{langchain}, a library for LLMs, to make requests to the GPT-4o model. With this setup, we utilized the LangChain framework to make subsequent requests to the GPT-4o model. Finally, we generate output code in the three languages we specified. Our dataset\footnote{https://github.com/llmForCapec/CAPECDatasetsLLM} is comprised of vulnerable code examples in Java, JavaScript, and Python programming languages. We selected these languages as they are the most widely used programming languages by developers \cite{proglang}.

To ensure the quality and utility of our dataset, we conducted a comprehensive evaluation focusing on the consistency of LLM code generation, relevance of the generated code to CAPEC, and readability of the code. While GPT-4o, Llama, and Claude were all used in this research, GPT-4o results were used for evaluation. Manual evaluations by independent reviewers confirmed that the generated code snippets closely align with the corresponding CAPEC and CWE descriptions, with strong inter-rater agreement. Furthermore, the evaluators rated the code samples for clarity and readability, giving them consistently high scores. These results highlight the clarity and structured nature of the code, making it a valuable resource for enhancing understanding and practical learning in cybersecurity education.

\begin{table}[h]
\centering
  \caption{Code Availability for CAPEC and CWE descriptions.}
  \label{tab:stats}
  \begin{tabular}{|c|c|c|c|c|}
    \hline
    Total CAPECs & Any Language & Java & JavaScript & Python\\
    \hline
    615 & \( \approx 61\% \) & \( \approx 25\% \) & \( \approx 0.6\% \) & \( \approx 3\% \) \\
    \hline
    \hline
    Total CWEs & Any Language & Java & JavaScript & Python\\
    \hline
    963 & \( \approx 53\% \) & \( \approx 16\% \) & \( \approx 0.5\% \) & \( \approx 2\% \)\\
    \hline
  \end{tabular}
  \vspace{-15pt}
\end{table}

\section{Related Work}

Deep learning techniques have also been applied to software vulnerability detection, introducing significant innovations. Large datasets like Big-Vul \cite{fan2020ac} and Common Architectural Weakness Enumeration (CAWE) \cite{santos2019empirical} have been developed to study code vulnerabilities and architectural weaknesses. Frameworks such as VulDeePecker \cite{li2018vuldeepecker}, SySeVR \cite{li2021sysevr}, and DiverseVul \cite{chen2023diversevul} show the high accuracy of deep learning methods in detecting vulnerabilities. Despite these advancements, challenges remain in applying deep learning to real-world vulnerability detection \cite{chakraborty2021deep}. A key issue is the reliance on pre-existing labeled datasets, which may not adequately represent emerging attack patterns or vulnerabilities.

The exploration of LLMs in software engineering and vulnerability detection is a growing area of research. Zhou et al. \cite{zhou2024large1} conducted a comprehensive literature review that categorizes various adaptation techniques in vulnerability detection and mitigation. Sun et al. \cite{sun2024llm4vuln} proposed a unified framework called LLM4Vuln, designed to enhance vulnerability reasoning. Additionally, analyses of LLMs' architecture, evolution, and ethical considerations across sectors such as education and healthcare provide a broad perspective on their applications \cite{bharathi2024analysis}. Several studies discuss the integration of LLMs into different aspects of the software development process, highlighting both the potential and challenges of this technology \cite{belzner2023large, fan2023large}. Contemporary research emphasizes thorough investigations into the effectiveness and challenges of LLMs in software security tasks. GPT-3.5 and GPT-4 can achieve competitive results with appropriate prompts \cite{zhou2024large2}. Zhang et al. \cite{zhang2024prompt} studied how ChatGPT detects software vulnerabilities with different prompts and found that ChatGPT identifies vulnerabilities in Java programs better than in C/C++ programs with a basic prompt but lacks a comprehensive understanding of vulnerabilities. 

Evaluations of various LLMs highlight their promise despite performance gaps when compared to static analysis tools \cite{purba2023software, pearce2023examining, noever2023can, fu2023chatgpt}. However, LLMs often require fine-tuning and domain-specific prompts to achieve optimal performance in specialized tasks \cite{ibra2024limtopic} \cite{ibra2024visual} like vulnerability detection, underscoring the importance of prompt engineering \cite{kojima2022large}. Some researchers have worked on providing code for specific CWEs, but there has not been any work in generating a code for CAPECs. We developed a methodology using GPT-4o and CAPEC \& CWE descriptions to generate example vulnerable code snippets. This approach not only addresses the scarcity of practical code examples in existing datasets but also provides contextually relevant outputs that align with theoretical descriptions of vulnerabilities. By utilizing semantic similarity techniques for CAPEC to CWE mapping and generating examples in widely used languages like Java, Python, and JavaScript, our work significantly contributes to the field. Our dataset is a valuable resource for researchers to not only understand vulnerabilities in code but also apply deep learning models to detect and mitigate vulnerabilities.

\section{METHODOLOGY}

\subsection{Mapping CAPEC and CWE}

The CAPEC and CWE datasets list various vulnerabilities and provide code examples. Sometimes, a CAPEC entry (a type of attack) doesn't have enough associated CWE entries (types of weaknesses). Because of this, we do not have sufficient textual information to provide GPT-4o, which could lead to inaccuracies in its generated results. To address this, we use SBERT\cite{reimers2019sentence}, a sentence transformer model, to find the most semantically similar CWEs for a given CAPEC entry. Specifically, we compare the CWE’s name, description, and extended description (if provided) to that of the CAPEC’s name and description. By using SBERT, we identify the most semantically related weaknesses for CAPECs.

Large Language Models (LLMs) are known to perform better when provided with detailed input prompts that include more context, examples, or demonstrations—a concept supported by research in few-shot learning\cite{brown2020language}. This approach leverages the ability of LLMs to use additional tokens effectively to generate higher-quality outputs. Inspired by this, we designed our methodology to augment CAPEC descriptions with related CWE descriptions to ensure that the input is both comprehensive and informative.

\begin{figure}
  \centering
  \includegraphics[width=1\columnwidth]{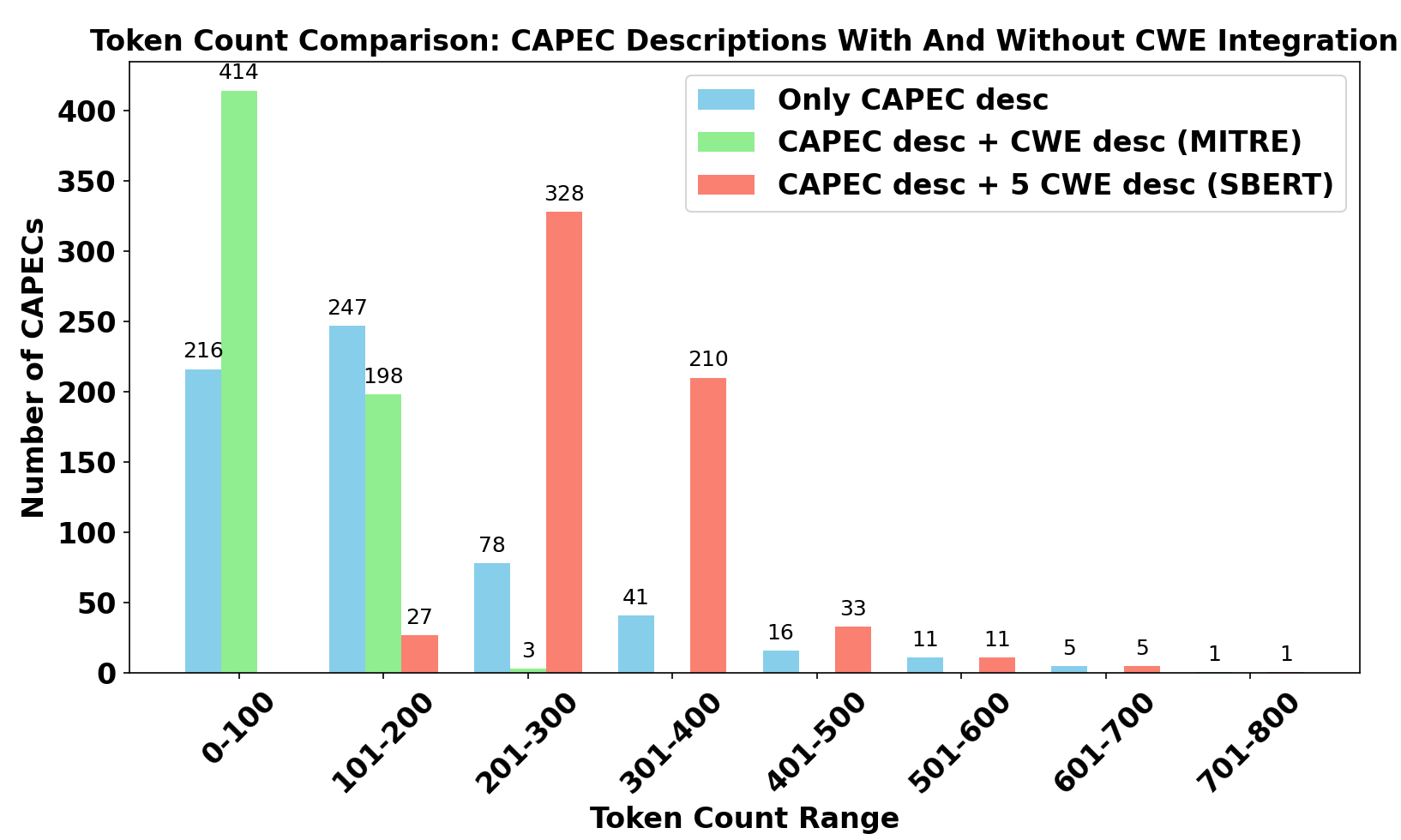}
  \caption{Token Count Comparison}
  \label{fig:token_stats}
  \vspace{-15pt}
\end{figure}

We analyzed the token counts for three input scenarios: (1) CAPEC descriptions alone, (2) CAPEC descriptions combined with all related CWE descriptions from MITRE, and (3) CAPEC descriptions combined with the top five most semantically similar CWE descriptions selected via SBERT. Figure \ref{fig:token_stats} illustrates how the token count increases across these scenarios, reflecting the additional context provided in the latter two cases. The increased input token count ensures that the LLM receives detailed and contextually enriched data, enhancing its ability to understand attack patterns and the associated weaknesses comprehensively. This integration of CAPEC and CWE descriptions addresses the limitations of CAPEC descriptions alone, enabling the generation of precise and contextually aligned outputs.

We consider the following two scenarios for mapping CAPECs to CWEs:

\textbf{Case 1 - CAPEC having 5 or more related CWEs:} If a CAPEC lists 5 or more related CWEs from the MITRE data, we include all of them. This is because we still want to use all the related weaknesses that MITRE provides. For example, in Table \ref{tab:capec_related_weaknesses}, CAPEC 1 has more than 5 related CWEs, so we consider all the CWEs as shown in Table \ref{tab:capec_top_weaknesses}.

\textbf{Case 2 - CAPEC having less than 5 related CWEs:} If a CAPEC lists fewer than 5 CWEs from the MITRE data, we still want to have at least 5 related CWEs for each CAPEC entry. To achieve this, we include all the CWEs listed by CAPEC from the MITRE data and then add additional CWEs based on their semantic similarity with the CAPEC. For example, in Table \ref{tab:capec_related_weaknesses}, CAPEC 18 has only one CWE (CWE 80) related to it, so we add the remaining four most semantically similar CWEs to make CAPEC 18 have at least 5 related CWEs as shown in Table \ref{tab:capec_top_weaknesses}.

We set a threshold of 5 CWEs to ensure a sufficient amount of textual data, including CAPEC descriptions and their associated CWE descriptions, is provided as input to GPT-4o. This threshold is adjustable and can be explored further to evaluate different values and determine the optimal setting

\begin{table*}

\centering
  \caption{Mapping CAPEC to their related CWEs using SBERT.}
  \label{tab:capec_related_weaknesses}
  
  \begin{tabular}{|c|c|c|}
  
    \hline
    \textbf{CAPEC ID} & \textbf{CWE IDs Related Weaknesses (from MITRE data)} & \textbf{CWE IDs Related Weaknesses (from SBERT)}\\
    \hline
    
     \multirow{2}{*}{1} & 276, 285, 434, 693, 732, 1191, 1193, 1220, 1297, & \multirow{2}{*}{ - } \\

& 1311, 1314, 1315, 1318, 1320, 1321, 1327 & \\
\hline

    2 & 645 & 645, 307, 521, 305, 1390\\
    
    \hline
    ... & ... & ...\\
    \hline
    
    18 & 80 & 82, 79, 80, 692, 81\\
    \hline

... & ... & ...\\
    \hline

    257 & NA & 217, 218, 534, 592, 533\\
    \hline

... & ... & ...\\
    \hline
    469 & 770, 772 & 488, 410, 770, 384, 535\\
    \hline
... & ... & ...\\
    \hline
    702 & 1296 & 1296, 1334, 1191, 1332, 1323\\
    \hline
  \end{tabular}
  \vspace{-10pt}
\end{table*}

\begin{table}
\centering
  \caption{Selecting Related CWEs (to feed into GPT-4o).}
  \label{tab:capec_top_weaknesses}
  \begin{tabular}{|c|c|}
  \hline
    \textbf{CAPEC ID} & \textbf{Selected CWE IDs Related Weaknesses} \\
    
    \hline
    \multirow{2}{*}{1} & 276, 285, 434, 693, 732, 1191, 1193, 1220, 1297, \\
 & 1311, 1314, 1315, 1318, 1320, 1321, 1327 \\
    \hline
    
    2 & 645, 307, 521, 305, 1390\\
    \hline
    ... & ...\\
    \hline
    18 & 80, 82, 79, 692, 81\\
    \hline
    ... & ...\\
    \hline
    257 & 217, 218, 534, 592, 533\\
    \hline
    ... & ...\\
    \hline
    469 & 770, 772, 488, 410, 384\\
    \hline
    ... & ...\\
    \hline
    702 & 1296, 1334, 1191, 1332\\
    \hline
  \end{tabular}
  \vspace{-10pt}
\end{table}

\subsection{Prompting GPT-4o for Code Generation}

We make use of GPT-4o, a state-of-the-art language model, to generate illustrative code snippets for each CAPEC. The process begins with collecting textual data from MITRE and using SBERT to enhance our CAPEC description data by finding additional related CWEs for each CAPEC. To facilitate this process, we employ LangChain, which serves as an interface between our CAPEC description dataset and GPT-4o. LangChain ingests the textual data for the CAPEC and CWEs, and communicates our crafted prompt to GPT-4o. It also ensures that the output is returned in our desired format.

In our approach, each CAPEC, along with its most closely related CWEs, is fed into GPT-4o. This input serves as the basis for generating a concise code snippet and a corresponding description of the code’s functionality. However, before this can occur, we must carefully craft a prompt that clearly communicates our desired output to the model. The prompt is an important artifact in getting better results \cite{zhang2024prompt}. The precision of the generated outcomes improves with the specificity of the instructions provided to the model \cite{kojima2022large, mungoli2023exploring}.

\begin{figure}[h]
    \centering
    \begin{tikzpicture}
        \node[draw, rectangle, rounded corners, fill=yellow!20, text width=0.5\textwidth, inner sep=5pt, scale=0.9] (box) {
            \begin{minipage}{\textwidth}
                \small
                \textbf{prompt} = \texttt{\textquotesingle\textquotesingle\textquotesingle} \\
                You are an expert in identifying and generating vulnerable code based on a given description. Given the CAPEC and related CWEs: [insert\_capec\_cwes], generate a [insert\_programming\_language] code snippet that embodies the main idea of the CAPEC, using the related CWEs for additional context. The code should be concise and represent the main point of the CAPEC. Also, provide a brief explanation of the code's functionality and the idea it represents. Respond in JSON format with `code\_snippet' and `description' keys.
                \texttt{\textquotesingle\textquotesingle\textquotesingle}
            \end{minipage}
        };
    \end{tikzpicture}
    \caption{Prompt for Code Generation}
    \label{fig:prompt_figure}
     \vspace{-8pt}
\end{figure}

We used the prompt as shown in Figure \ref{fig:prompt_figure}. Having a prompt structured in this manner generates more accurate results \cite{lyu2023faithful,zhou2024large2}.  Once the prompt is prepared, we populate the placeholder [insert\_capec\_cwes] with the relevant textual data (descriptions, names, extended descriptions) to provide context for the CAPEC and its most related CWEs. The placeholder [insert\_programming\_language] is populated with the specific programming language. Providing well-structured and detailed prompts is crucial for generating accurate code, as GPT-4o's results depend heavily on the quality of the input it receives.

To ensure the model remains focused on the task at hand, we explicitly instruct it to generate concise code that encapsulates the main point of the CAPEC. Without this directive, the model might fail to capture the essence of the attack pattern and instead produce code that is more aligned with the weaknesses, thereby missing the primary objective of the attack pattern. We also prompt the model to articulate its thought process, providing a description of the code’s functionality and the underlying idea it represents. The final part of the prompt specifies the desired response format.

Upon completion of the model’s processing, we obtain two key outputs. The first is a code snippet that encapsulates the essence of the CAPEC and its most related CWEs. This snippet serves as a tangible representation of the vulnerabilities described. The second output is a comprehensive description that elucidates the functionality of the code snippet. This description provides an overview of what the code is doing, and how it relates to the associated CAPEC and CWE. Additionally, the output code is annotated with inline comments. These comments offer insights into the functionality of each line of code and indicate the specific CWE it represents. Together, these outputs provide a detailed and practical understanding of each CAPEC.

\section{Results}
Our research with the GPT-4o model yields comprehensive results, including generated code snippets in JavaScript, Python, and Java for all CAPECs and their associated CWEs, along with detailed descriptions of the code’s functionality. This output not only deepens our understanding of the CAPECs and CWEs but also serves as a valuable resource for future model training, thereby enhancing our ability to understand and detect code vulnerabilities.

\begin{figure}
  \centering
  \includegraphics[width=1.03\columnwidth]{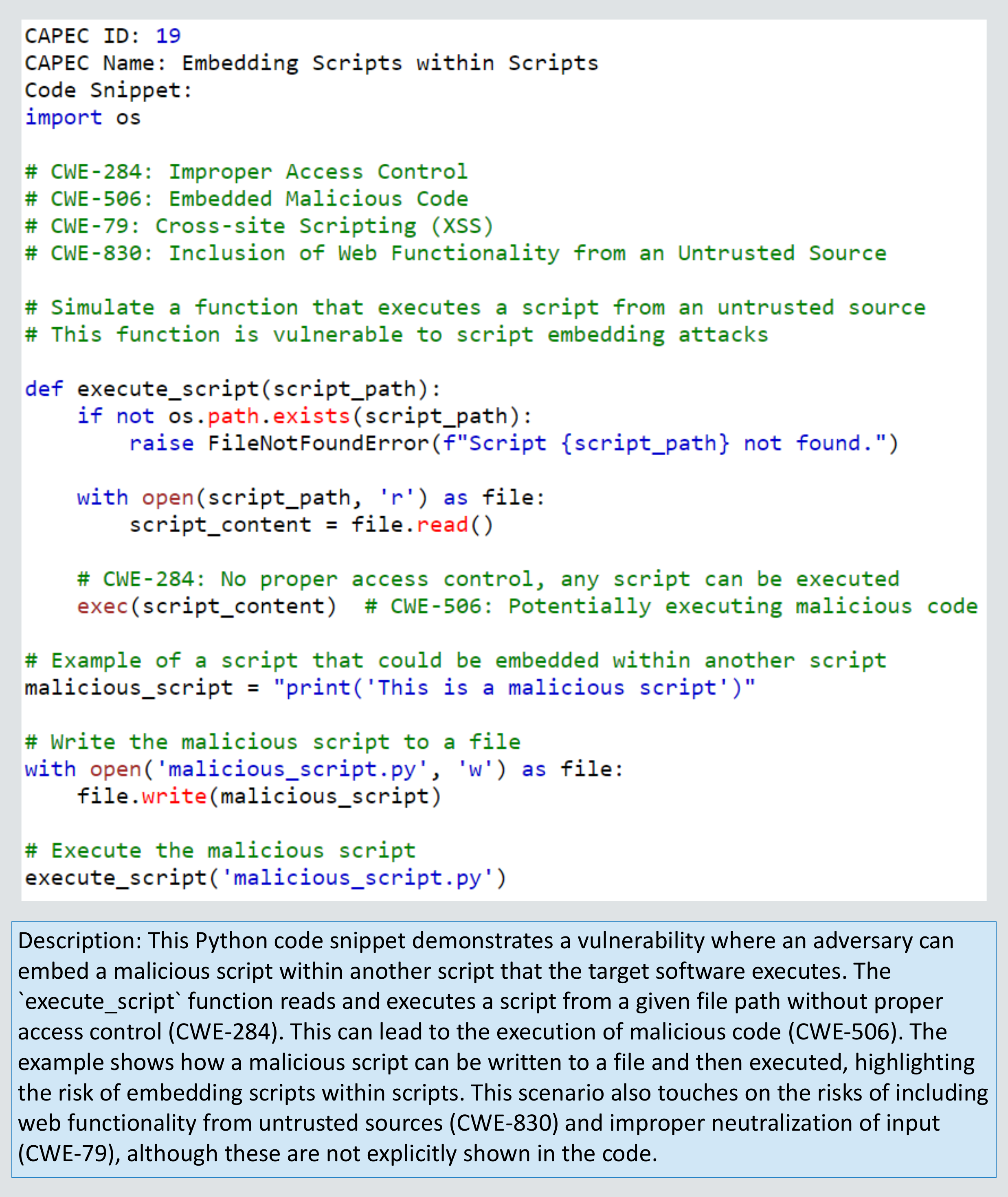}
  \caption{GPT-4o output for CAPEC 19: Embedding Scripts within Scripts.}
  \label{fig:code_snippet}
  \vspace{-15pt}
\end{figure}

Figure \ref{fig:code_snippet} shows an example from the generated dataset. It illustrates the output of GPT-4o for \textit{CAPEC-19: Embedding Scripts within Scripts} in Python language. GPT-4o creates a function named ‘execute\_script’. This function reads and executes another script from a file. The key operation here is that it takes a file path as input, reads the content of the file, and then executes it. However, the critical issue with this function is the lack of proper access control. It does not verify or restrict which scripts can be executed, potentially leading to the execution of any script, including those that are malicious.

The documentation comments provided in the output given by GPT-4o help in understanding the code and its implications. These comments highlight the specific CWEs that are relevant to this code snippet. They provide context about the potential vulnerabilities that this code might introduce, such as \textit{Improper Access Control (CWE-284)} and \textit{Embedded Malicious Code (CWE-506)}. In addition to the code and comments, a detailed description is provided. This description explains the functionality of the code and the potential security risks it represents. It serves as a valuable guide for Python programmers to understand the code, its operation, and its implications. Together, the code, comments, and description provide a comprehensive understanding of the CAPEC and its representation in the code. They show the thought process of the model and help users understand the ‘why’, ‘how’, and ‘what’ of the CAPEC. This multi-faceted approach is crucial for fully understanding and learning from the code example.

\section{Evaluation}

The evaluation of the generated dataset is divided into the following key aspects:

\begin{enumerate}[label=\Alph*)]
    \item Evaluation of the generated code
    \begin{enumerate}[label=\arabic*)]
        \item Compilability of the generated code
        \item Relevance of the generated code to CAPEC
        \item Code Readability
    \end{enumerate}
    \item Consistency of GPT-4o output
    \item Consistency among different LLMs' outputs
    
\end{enumerate}

\subsection{Evaluation of the generated code}
For the evaluation of code, some metrics are available, such as unit testing and Pass@k metric \cite{chen2021evaluating}. While these metrics are helpful, they are often limited to specific contexts or tasks. A universally accepted, holistic metric for evaluating all aspects of LLM-generated code is still an open research area. As a result, we rely on manual evaluation to review the generated code. Given that this is a labor-intensive process, we focus our evaluation solely on the GPT-4o-generated datasets. From these datasets, we randomly select 30 code snippets per language for assessment. As shown in Table \ref{tab:differentLLMs}, Llama and Claude produce similar code outputs, allowing us to reasonably assume that manual assessment of their datasets would yield comparable results.

\subsubsection{Compilability of the generated code}

We randomly selected 30 code snippets for each programming language from the five datasets generated by GPT-4o and tested their ability to compile. Table \ref{tab:compilation_table} summarizes the number of samples that were successfully compiled. The results show that approximately 90\% of the code snippets compiled successfully across most datasets, with Python exhibiting the highest compilation success rate, followed by Java and then JavaScript.

\begin{table}[H]
\vspace{-10pt}
\centering
\caption{Code Compilation for 30 Random Samples}
\label{tab:compilation_table}
\resizebox{\columnwidth}{!}{%
\begin{tabular}{|c|c|c|c|c|c|}
\hline
Language & Dataset 1 & Dataset 2 & Dataset 3 & Dataset 4 & Dataset 5 \\
\hline
Java & 28 & 29 & 26 & 29 & 27 \\
\hline
Python & 27 & 29 & 30 & 29 & 29 \\
\hline
JavaScript & 24 & 24 & 27 & 27 & 27 \\
\hline
\end{tabular}%
}
\vspace{-7pt}
\end{table}

\subsubsection{Relevance of the generated code to CAPEC}
To evaluate the relevance of the generated code to CAPEC, we conducted a user study using the same set of 30 randomly selected code snippets per programming language that were previously tested for compilability. These samples were drawn from two datasets generated by GPT-4o. To keep the user study straightforward and manageable for the evaluators, we limited the evaluation to two datasets. The samples were independently reviewed by four researchers, all of whom are Computer Science Ph.D students with strong coding expertise, ensuring the reliability of the assessment. While one of the evaluators has since joined as a co-author of this paper, their evaluation was conducted independently, prior to their authorship, to maintain objectivity and minimize any potential bias.

The evaluation involved reviewing each code snippet and its associated CAPEC description to assess their coherence, logical validity, and accuracy in representing the relevant attack pattern. If a snippet and its description met these criteria, the researcher marked it as ``yes"; otherwise, it was marked as ``no". This systematic approach allowed us to objectively assess the quality and relevance of the generated code snippets and their corresponding descriptions. Table \ref{tab:accuravyeval} presents the number of sample code snippets evaluated as relevant to CAPEC by the researchers. We see that three of the evaluators agree that 100\% of the code snippets are relevant to the CAPEC descriptions.

In order to quantify the degree of agreement between the evaluators, we need an inter-rater agreement metric. The choice of this metric is crucial for accurately assessing reliability, particularly in datasets with specific characteristics. In this study, we utilized Average Pairwise Agreement (APA) due to its suitability for datasets with high agreement levels, small sample sizes, and imbalanced categories. We found that the APA for Java, Python, and JavaScript datasets were 95.83\%, 93.33\%, and 96.67\%, respectively. These APA values demonstrate strong consensus among raters, with only a few isolated disagreements. Unlike Fleiss' Kappa and Krippendorff's Alpha, which penalize isolated disagreements disproportionately and adjust for chance agreement, APA provides a direct and intuitive measure of actual agreement, offering meaningful results in contexts with high agreement \cite{lombard2002content}. Additionally, APA is less sensitive to small sample sizes, avoiding the sharp score reductions seen with Kappa metrics, which can misrepresent reliability in limited datasets \cite{mchugh2012interrater}. Moreover, the imbalanced nature of these datasets, where ``yes" responses dominate, makes APA particularly advantageous, as it avoids biases inherent in chance-adjusted measures that can be skewed in favor of minority categories \cite{stemler2019comparison}.

\begin{table}[h!]
\caption{Code Relevance to CAPEC of 30 Random Samples and Code Readability Rating}
    \centering
    \resizebox{\columnwidth}{!}{%
    \begin{tabular}{|c|c|c|c|c|c|}
        \hline  
        \multicolumn{2}{|c|}{} & \multicolumn{2}{c|}{Relevance to CAPEC} & \multicolumn{2}{c|}{Code Readability(1-5)} \\
        \hline
        \multicolumn{1}{|c|}{} & Language & Dataset 1 & Dataset 2 & Dataset 1 & Dataset 2 \\
        \hline
                \multirow{3}{*}{Evaluator 1} & Java & 30 & 30 & 4 & 4 \\
        \cline{2-6}
                                    & Python & 30 & 30 & 5 & 5 \\
        \cline{2-6}
                                    & JavaScript & 30 & 30 & 5 & 5 \\
        \hline
        \multirow{3}{*}{Evaluator 2} & Java & 27 & 28 & 4 & 4 \\
        \cline{2-6}
                                    & Python & 27 & 27 & 5 & 5 \\
        \cline{2-6}
                                    & JavaScript & 28 & 28 & 3 & 3 \\
        \hline
        \multirow{3}{*}{Evaluator 3} & Java & 30 & 30 & 5 & 5 \\
        \cline{2-6}
                                    & Python & 30 & 30 & 5 & 5 \\
        \cline{2-6}
                                    & JavaScript & 30 & 30 & 5 & 5 \\
        \hline
        \multirow{3}{*}{Evaluator 4} & Java & 30 & 30 & 4 & 4 \\
        \cline{2-6}
                                    & Python & 30 & 30 & 4 & 4 \\
        \cline{2-6}
                                    & JavaScript & 30 & 30 & 4 & 4 \\
        \hline
    \end{tabular}%
    }
    \label{tab:accuravyeval}
\end{table}

\subsubsection{Code Readability}
In addition to evaluating the code's relevance to CAPEC, we asked the evaluators to rate the code's clarity and ease of understanding on a scale of 1 to 5, with 5 indicating the highest level of clarity and readability. The ratings, as shown in the table \ref{tab:accuravyeval}, highlight a high level of consistency among the evaluators, with scores predominantly in the range of 4 to 5 across all programming languages and datasets. Notably, the code generated in Python was rated as the most readable, consistently achieving higher scores compared to Java and JavaScript. This suggests that the Python code samples were particularly well-written and structured, making them easier to understand. The minor differences in the ratings likely reflect subjective evaluator preferences or differing familiarity with the programming languages. Overall, the narrow range of scores demonstrates strong inter-rater agreement, underscoring the clarity of the provided code and the effectiveness of the coding practices employed.

\subsection{Consistency of GPT-4o output}

We conduct this evaluation to determine whether GPT-4o is reliable in generating consistent code outputs rather than producing random variations when given the same input. To assess the consistency of GPT-4o in generating similar code, we generated five distinct datasets under a consistent experimental design. These datasets were then compared to evaluate their similarity, providing insights into the model's consistency. For measuring the similarity between the generated code snippets, we utilize CodeBERT\cite{feng2020codebert}, a pre-trained language model specifically designed for programming languages and natural language processing tasks. Using CodeBERT, we calculate the cosine similarity between the generated code in multiple datasets. As shown in Table \ref{tab:datasets}, the results show high similarity scores: Java and Python exceed a cosine similarity of 0.99, while JavaScript achieves over 0.98. These high similarity scores lead us to conclude that GPT-4o consistently produces the same output under identical conditions.

\begin{table}[h!]
    \centering
    \scriptsize 
    \caption{Cosine Similarity Among Code Snippets in datasets generated by GPT-4o.}
    \label{tab:datasets}
    
    \begin{tabularx}{\columnwidth}{|l|X|X|X|X|X|}
        \hline
        \multicolumn{6}{|c|}{\textbf{Java}} \\ \hline
        & \textbf{Dataset 1} & \textbf{Dataset 2} & \textbf{Dataset 3} & \textbf{Dataset 4} & \textbf{Dataset 5} \\ \hline
        \textbf{Dataset 1} & - & 0.9938 & 0.9938 & 0.9938 & 0.9939 \\ \hline
        \textbf{Dataset 2} & 0.9938 & - & 0.9933 & 0.9934 & 0.9935 \\ \hline
        \textbf{Dataset 3} & 0.9938 & 0.9933 & - & 0.9933 & 0.9934 \\ \hline
        \textbf{Dataset 4} & 0.9938 & 0.9934 & 0.9933 & - & 0.9935 \\ \hline
        \textbf{Dataset 5} & 0.9939 & 0.9935 & 0.9934 & 0.9935 & - \\ \hline
    \end{tabularx}

    \begin{tabularx}{\columnwidth}{|l|X|X|X|X|X|}
        \hline
        \multicolumn{6}{|c|}{\textbf{Python}} \\ \hline
        & \textbf{Dataset 1} & \textbf{Dataset 2} & \textbf{Dataset 3} & \textbf{Dataset 4} & \textbf{Dataset 5} \\ \hline
        \textbf{Dataset 1} & - & 0.9904 & 0.9903 & 0.9904 & 0.9903 \\ \hline
        \textbf{Dataset 2} & 0.9904 & - & 0.9904 & 0.9904 & 0.9904 \\ \hline
        \textbf{Dataset 3} & 0.9903 & 0.9904 & - & 0.9903 & 0.9903 \\ \hline
        \textbf{Dataset 4} & 0.9904 & 0.9904 & 0.9903 & - & 0.9903 \\ \hline
        \textbf{Dataset 5} & 0.9903 & 0.9904 & 0.9903 & 0.9903 & - \\ \hline
    \end{tabularx}

    \begin{tabularx}{\columnwidth}{|l|X|X|X|X|X|}
        \hline
        \multicolumn{6}{|c|}{\textbf{JavaScript}} \\ \hline
        & \textbf{Dataset 1} & \textbf{Dataset 2} & \textbf{Dataset 3} & \textbf{Dataset 4} & \textbf{Dataset 5} \\ \hline
        \textbf{Dataset 1} & - & 0.9891 & 0.9891 & 0.9890 & 0.9890 \\ \hline
        \textbf{Dataset 2} & 0.9891 & - & 0.9896 & 0.9896 & 0.9896 \\ \hline
        \textbf{Dataset 3} & 0.9891 & 0.9896 & - & 0.9896 & 0.9896 \\ \hline
        \textbf{Dataset 4} & 0.9890 & 0.9896 & 0.9896 & - & 0.9895 \\ \hline
        \textbf{Dataset 5} & 0.9890 & 0.9896 & 0.9896 & 0.9895 & - \\ \hline
    \end{tabularx}
    \vspace{-10pt}
\end{table}

\subsection{Consistency among different LLMs' outputs}

To evaluate the consistency of our approach across different LLMs, we examined how similar the outputs are when using various models. For this purpose, we repeated the same experiment with multiple LLMs. In addition to GPT-4o, the following models were utilized:

\begin{itemize} 
\item \textbf{Llama}: Version 3 with 70 billion parameters and a context window of 8,192 tokens, which is sufficient for our experiment design.

\item \textbf{Claude}: Claude-3-5-sonnet was selected for its advanced contextual understanding and ability to generate precise and coherent outputs, making it well-suited to the requirements of our experiment.
\end{itemize}

\begin{table*}[!t] 
    \centering
    \scriptsize
    \caption{Cosine Similarity of datasets generated by GPT-4o, Llama, and Claude for different programming languages.}
    \label{tab:differentLLMs}
    
    \begin{tabularx}{\textwidth}{|l|X|X|X|X|X|X|}
        \hline
        \multicolumn{7}{|c|}{\textbf{Java}} \\ \hline
        & \textbf{GPT-4o Dataset 1} & \textbf{GPT-4o Dataset 2} & \textbf{Llama Dataset 1} & \textbf{Llama Dataset 2} & \textbf{Claude Dataset 1} & \textbf{Claude Dataset 2} \\ \hline
        \textbf{GPT-4o Dataset 1} & - & 0.9938 & 0.9916 & 0.9915 & 0.9936 & 0.9936 \\ \hline
        \textbf{GPT-4o Dataset 2} & 0.9938 & - & 0.9912 & 0.9911 & 0.9931 & 0.9931 \\ \hline
        \textbf{Llama Dataset 1} & 0.9916 & 0.9912 & - & 0.9894 & 0.9910 & 0.9911 \\ \hline
        \textbf{Llama Dataset 2} & 0.9915 & 0.9911 & 0.9894 & - & 0.9910 & 0.9910 \\ \hline
        \textbf{Claude Dataset 1} & 0.9936 & 0.9931 & 0.9910 & 0.9910 & - & 0.9944 \\ \hline
        \textbf{Claude Dataset 2} & 0.9936 & 0.9931 & 0.9911 & 0.9910 & 0.9944 & - \\ \hline
    \end{tabularx}
    
    \begin{tabularx}{\textwidth}{|l|X|X|X|X|X|X|}
        \hline
        \multicolumn{7}{|c|}{\textbf{Python}} \\ \hline
        & \textbf{GPT-4o Dataset 1} & \textbf{GPT-4o Dataset 2} & \textbf{Llama Dataset 1} & \textbf{Llama Dataset 2} & \textbf{Claude Dataset 1} & \textbf{Claude Dataset 2} \\ \hline
        \textbf{GPT-4o Dataset 1} & - & 0.9904 & 0.9887 & 0.9887 & 0.9883 & 0.9876 \\ \hline
        \textbf{GPT-4o Dataset 2} & 0.9904 & - & 0.9888 & 0.9887 & 0.9883 & 0.9876 \\ \hline
        \textbf{Llama Dataset 1} & 0.9887 & 0.9888 & - & 0.9874 & 0.9863 & 0.9856 \\ \hline
        \textbf{Llama Dataset 2} & 0.9887 & 0.9887 & 0.9874 & - & 0.9862 & 0.9855 \\ \hline
        \textbf{Claude Dataset 1} & 0.9883 & 0.9883 & 0.9863 & 0.9862 & - & 0.9874 \\ \hline
        \textbf{Claude Dataset 2} & 0.9876 & 0.9876 & 0.9856 & 0.9855 & 0.9874 & - \\ \hline
    \end{tabularx}
    
    \begin{tabularx}{\textwidth}{|l|X|X|X|X|X|X|}
        \hline
        \multicolumn{7}{|c|}{\textbf{JavaScript}} \\ \hline
        & \textbf{GPT-4o Dataset 1} & \textbf{GPT-4o Dataset 2} & \textbf{Llama Dataset 1} & \textbf{Llama Dataset 2} & \textbf{Claude Dataset 1} & \textbf{Claude Dataset 2} \\ \hline
        \textbf{GPT-4o Dataset 1} & - & 0.9891 & 0.9864 & 0.9863 & 0.9856 & 0.9862 \\ \hline
        \textbf{GPT-4o Dataset 2} & 0.9891 & - & 0.9869 & 0.9868 & 0.9861 & 0.9868 \\ \hline
        \textbf{Llama Dataset 1} & 0.9864 & 0.9869 & - & 0.9845 & 0.9835 & 0.9841 \\ \hline
        \textbf{Llama Dataset 2} & 0.9863 & 0.9868 & 0.9845 & - & 0.9834 & 0.9840 \\ \hline
        \textbf{Claude Dataset 1} & 0.9856 & 0.9861 & 0.9835 & 0.9834 & - & 0.9853 \\ \hline
        \textbf{Claude Dataset 2} & 0.9862 & 0.9868 & 0.9841 & 0.9840 & 0.9853 & - \\ \hline
    \end{tabularx}
    \vspace{-15pt}
\end{table*}

For all three LLMs, we used the same prompt to generate two datasets for each programming language. The generated code was evaluated for similarity using the CodeBERT model, which calculates cosine similarity scores. Table \ref{tab:differentLLMs} presents the cosine similarity scores for the datasets generated by the three LLMs. We see that the cosine similarity values show strong alignment across datasets generated by GPT-4o, Llama, and Claude. Java datasets exhibit the highest similarity, while JavaScript datasets display slightly more variation.

\section{Discussion}

\subsection{Interpretation of Results and Implications}
The results demonstrate the efficacy of using GPT-4o, Llama, and Claude-3-5-sonnet for generating illustrative code snippets aligned with CAPEC and CWE descriptions. These models consistently produced high-quality outputs with measurable accuracy and applicability for cybersecurity research and training. The generated snippets in Java, Python, and JavaScript provide actionable insights into vulnerabilities, bridging the gap between theoretical descriptions and practical applications.

The dataset has several implications. It offers a robust foundation for training machine learning models to identify and mitigate vulnerabilities, serves as an educational resource for teaching secure coding practices, and supports automated tools for enhanced code analysis. However, users are advised to validate the relevance of the generated code snippets to ensure their alignment with the respective CAPEC descriptions, particularly for non-software-related CAPECs.

\subsection{Limitations and Future Work}
Despite its contributions, this study has certain limitations. Some CAPEC entries, such as CAPEC-391 (Bypassing Physical Locks) and CAPEC-547 (Physical Destruction of Devices), are unrelated to software but still had code generated for them.  CAPEC does not provide explicit classifications separating software-related attack patterns from others. This highlights the need for a pre-filtering step to classify CAPECs into software-related and non-software-related categories to improve dataset applicability.

We chose five related CWEs to provide enough context for GPT-4o without overwhelming it. Using fewer CWEs could make the LLM focus more on the CAPEC, while using more might dilute its focus. Future work can explore adjusting this number to balance context and relevance. The choice of Claude-3-5-sonnet was based on its accessibility and suitability for the study, though more advanced versions, such as Claude-3 Opus, optimized for software code generation, may yield improved results. Expanding the dataset to include additional programming languages, such as C++ or Go, could also enhance its versatility. 

Future work could focus on addressing these limitations by incorporating a classification mechanism for CAPEC entries, exploring advanced LLMs like Claude-3 Opus, and integrating the dataset with real-world vulnerability scanning tools. These enhancements would further solidify the utility of this resource in cybersecurity research and practice.

\section{Conclusion}

This research presents a robust methodology for generating practical, illustrative code examples of cybersecurity vulnerabilities using the GPT-4o model and CAPEC-CWE mappings. By addressing the scarcity of accessible code examples in existing datasets, this study contributes a unique, contextually rich resource that deepens understanding of software vulnerabilities and enhances the ability to detect and mitigate them. The generated snippets, evaluated for accuracy, readability, and applicability across Java, JavaScript, and Python, provide a reliable foundation for researchers, developers, and educators to study attack patterns and weaknesses in greater detail.

This work underscores the efficacy of GPT-4o in generating contextually relevant and accurate code representations, bridging the gap between theoretical descriptions and real-world applications of vulnerabilities. The dataset not only serves as a valuable tool for advancing machine learning-based vulnerability detection but also fosters secure coding practices and hands-on learning in cybersecurity. The resource contributes significantly to the broader field of software security by providing a practical and powerful means of studying and preventing vulnerabilities.

Future directions may include expanding the dataset to encompass additional programming languages and exploring its integration into real-world vulnerability assessment and remediation tools. These advancements have the potential to further strengthen the study and prevention of cybersecurity threats, marking a significant leap forward in the field.

\textbf{Acknowledgments:} We would like to thank the researchers who participated in the user study for the evaluation of results.

\end{document}